\documentclass{optica-article}

\journal{opticajournal} 

\articletype{Research Article}
\usepackage{subfigure}
\usepackage{subcaption}
\usepackage[justification=raggedright,singlelinecheck=false]{caption} 
\hypersetup{
  colorlinks=true,
  linkcolor=blue,
  citecolor=blue,
  urlcolor=blue
}
\renewcommand{\thesubfigure}{(\alph{subfigure})}
\usepackage{lineno}
\usepackage{bm}
\usepackage[normalem]{ulem}
\usepackage{graphicx}
\begin{document}

\title{Observation of an Optical Spring in a\\  Robustly Controlled Signal-Recycled \\Michelson Interferometer}

\author{Kaido Suzuki,\authormark{1,*}
Ken-ichi Harada,\authormark{2}
Ryo Iden,\authormark{1}
Sotatsu Otabe,\authormark{3,4}
Kentaro Komori,\authormark{3,5}
Yuta Michimura,\authormark{5,6}\\
and 
Kentaro Somiya\authormark{1}}

\address{\authormark{1}Department of Physics, Institute of Science Tokyo, Meguro, Tokyo 152-8550, Japan\\
\authormark{2}Green Computing Systems Research Organization, Waseda University, 27, Waseda-cho, Shinjuku-ku, Tokyo 162-0042, Japan\\
\authormark{3}Department of Physics, University of Tokyo, Bunkyo, Tokyo 113-0033, Japan\\
\authormark{4}Institute of Integrated Research, Institute of Science Tokyo, Yokohama, Kanagawa 226-8503, Japan\\
\authormark{5}Research Center for the Early Universe (RESCEU), Graduate School of Science,
University of Tokyo, Bunkyo, Tokyo 113-0033, Japan\\
\authormark{6}Kavli Institute for the Physics and Mathematics of the Universe (Kavli IPMU), WPI, UTIAS, University of Tokyo, Kashiwa, Chiba 277-8568, Japan\\
}

\email{\authormark{*}ksuzuki@gw.phys.sci.isct.ac.jp} 


\begin{abstract*}
We present, to the best of our knowledge, the first demonstration of an optical spring in a signal-recycled Michelson interferometer without arm cavities. 
The setup utilizes multiple laser fields to achieve stable and precisely controllable detuning and incorporates a compliant suspension system comprising two double-spiral springs.
These results support optical spring-enhanced interferometers with intracavity amplification, offering a promising avenue for improving high-frequency sensitivity in future gravitational-wave detectors.
\end{abstract*}

\section{Introduction}
Since the first detection of gravitational waves in 2015 \cite{150914}, numerous subsequent detections have been reported, primarily originating from mergers involving black holes and neutron stars \cite{alarts}.
Despite improved detector sensitivity, gravitational waves above 1\,kHz have not been observed due to the insufficient sensitivity of current detectors. 
High-frequency signals, particularly those emitted during the post-merger phase of binary neutron star collisions, are key to understanding matter under extreme conditions and probing the interiors of neutron stars \cite{BNS}. 
This highlights the need for enhanced sensitivity in the kilohertz band. 
A major limitation to high-frequency sensitivity is quantum shot noise, which dominates above several hundred hertz in second-generation gravitational-wave detectors such as Advanced LIGO~\cite{LIGO}, Advanced Virgo~\cite{virgo}, and KAGRA~\cite{KAGRA}. 
Although current detector technologies utilize squeezed vacuum injection to partially reduce this limitation \cite{LIGOsqz, LIGOfdsqz, virgofdsqz}, optomechanical interactions offer a complementary strategy. In such systems, radiation pressure couples to the mechanical suspension of the mirrors, creating an optical spring that alters the mechanical response of the detector \cite{cavoptmech}. This optical spring effect enhances sensitivity at the spring resonance frequency, and tuning the optical spring constant allows the resonance to shift to higher frequencies, thereby improving sensitivity in the kHz band.
The optical spring constant can be tuned by adjusting the detuning phase of the signal-recycling cavity (SRC) from resonance.
Its maximum achievable value is directly proportional to the laser power incident on the test mass \cite{osshread}. 
However, increasing laser power is practically constrained by adverse effects such as thermal lensing \cite{lensing} and parametric instabilities \cite{parainst}.

An alternative method to enhancing the optical spring constant involves incorporating an optical parametric amplifier (OPA) within the signal-recycling cavity \cite{somiya2016}. Optical parametric amplification enables the interferometer to surpass the Mizuno limit—a fundamental constraint on the product of sensitivity and bandwidth—by modifying its quantum noise response \cite{korobko2017}. With an intracavity OPA, substantial enhancement of the optical spring constant can be achieved without increasing the incident laser power.
The manipulation of mechanical responses via optical rigidity has been demonstrated in several experiments \cite{osshread,mani,anets}. In 2006, Miyakawa \textit{et al.} reported a significant observation: an optical spring with stiffness exceeding that of the mechanical suspension. This result was realized using a dual-recycled interferometer configuration with arm cavities, known as a resonant sideband extraction (RSE) interferometer\cite{osrse}. Around the same time, Corbitt \textit{et al.} demonstrated a high-stiffness optical spring using a single optical cavity \cite{oscorbit}. While most prior experiments utilized high-power cavities, Cripe \textit{et al.} successfully observed an optical spring in a hybrid setup combining Michelson and Sagnac interferometric configurations \cite{Cripe2018}.

More recently, improvements in optical spring performance have been demonstrated. Otabe \textit{et al.} introduced a nonlinear crystal for OPA into an optical cavity operated under phase-mismatched conditions, observing variations in the optical spring constant mediated by the optical Kerr effect \cite{otabe2024}.
Liu \textit{et al.} implemented an intracavity OPA under phase-matched conditions using a suspended membrane, demonstrating measurable changes in the optical spring constant \cite{Liu:25}. 
Despite these successful demonstrations, incorporating an OPA crystal into a high-power cavity environment poses significant thermal management challenges \cite{OtabePT}. 
Furthermore, our previous research has shown that optical spring enhancement due to intracavity signal amplification becomes more pronounced in interferometer configurations without arm cavities, owing to their reduced internal phase delay \cite{somiya2016}.
Thus, signal-recycled interferometers without arm cavities offer a particularly suitable platform for implementing optical parametric amplification.
The phase delay introduced by arm cavities could be mitigated by employing an additional signal-recycling mirror (SRM) that forms a critically coupled cavity with the input mirrors \cite{WLC}. 
An OPA could be inserted between the additional mirror and the main SRM. 
Such a scheme introduces further complexity into the control system but remains an option for increasing the circulating power. 

Observing an optical spring in interferometer configurations without arm cavities has presented significant challenges. 
The GEO600 gravitational-wave detector, which adopts this configuration, achieved notable sensitivity down to frequencies of tens of hertz during the mid-2000s and conducted extensive investigations of optical spring effects; however, these effects remained undetectable \cite{GEO_OS_attempt}. 
In this work, we overcome these challenges by employing a robust control scheme that combines multiple laser fields and a stable suspension system. 
While radio-frequency sidebands on the fundamental carrier, as demonstrated at the Caltech 40\,m interferometer \cite{Adhikari2008}, would couple to SRC length variations, 
we instead employed a green second-harmonic beam—commonly used for arm cavity acquisition in gravitational-wave detectors \cite{Mullavey:12}—to stably control the Michelson differential arm length independently of SRC detuning. 
For the suspension system, we constructed a suspension consisting of two double-spiral springs, which provide high compliance in the longitudinal degree of freedom while remaining rigid in pitch and yaw, and further introduced magnetic eddy-current damping to suppress excess motion.
These measures enabled stable operation and allowed us to continuously measure the transfer function over a wide range of detuning angles without losing lock.

This paper is organized as follows. 
In Section~\ref{sec:theory}, we introduce the theoretical framework for the optical spring in an SRMI. 
Section~\ref{sec:experiment} describes the experimental setup and control scheme, designed to overcome the challenges outlined above. 
In Section~\ref{sec:results}, we present the measurement results and compare them with theoretical predictions. 
Finally, Section~\ref{sec:outlook} discusses the implications of our results for future implementations, particularly in the context of intracavity OPA.

\section{Theory}\label{sec:theory}

\subsection{Optical spring constant}
\begin{figure}[htbp]
\centering\includegraphics[width=7cm]{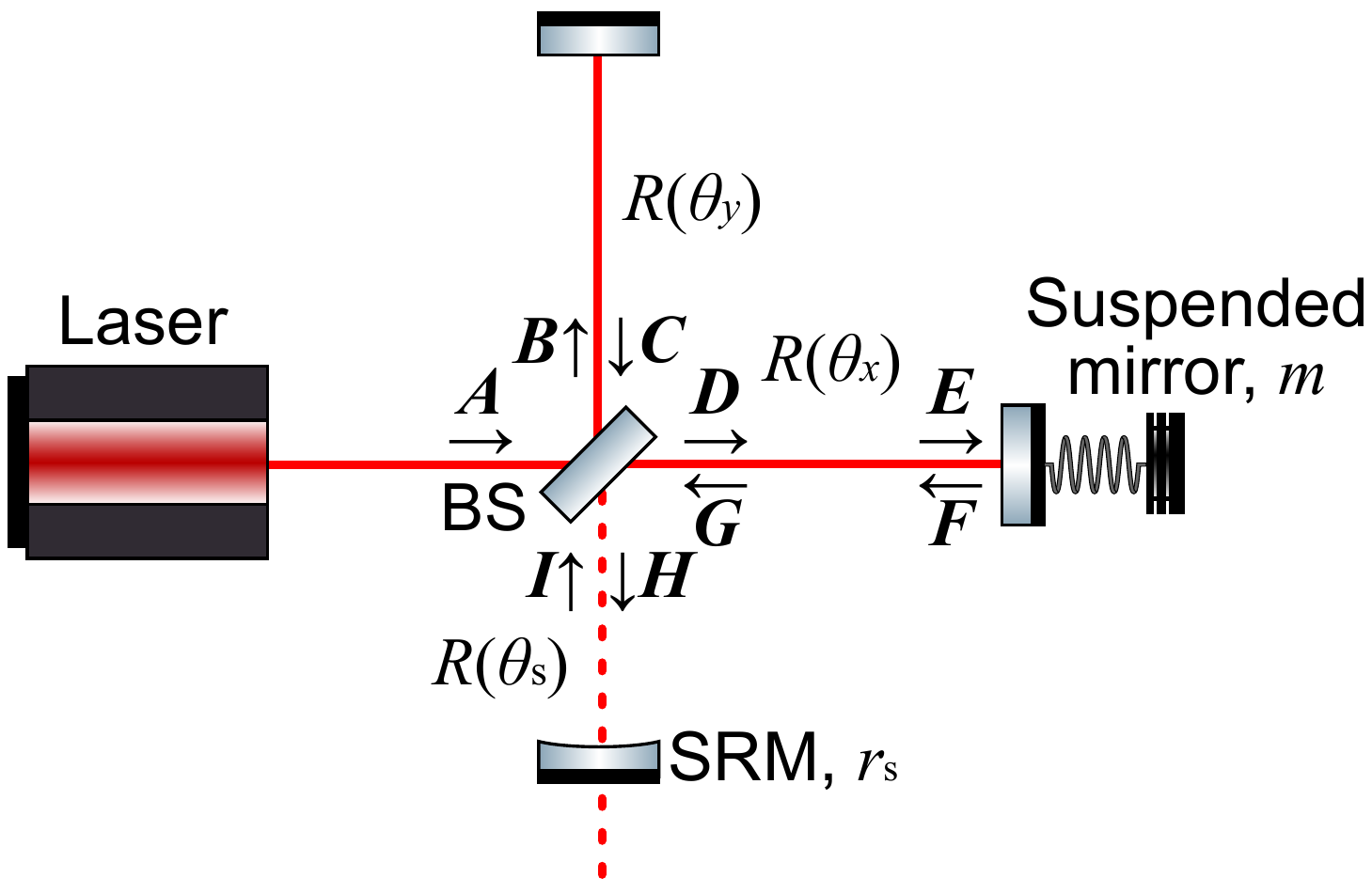}
\caption{Schematic of the SRMI incorporating a suspended mirror with mass $m$. The input electric field $\bm{A}$ from the laser source is split by a 50:50 beam splitter (BS), reflected by the end mirrors, and recombined at the beam splitter. The output field $\bm{H}$ is partially reflected by the SRM, with reflectivity $r_\mathrm{s}$, and undergoes additional interference within the signal-recycling cavity.}
\label{fig:SRMI_IO}
\end{figure}

To investigate the behavior of optical springs in the present system, the input–output relations \cite{IOrel} for an SRMI with a suspended end mirror are employed (see Fig.~\ref{fig:SRMI_IO}): 

\begin{align}
  \bm{B} &= \frac{1}{\sqrt{2}} \bm{A} + \frac{1}{\sqrt{2}} \bm{I}, \quad   
  \bm{D} = \frac{1}{\sqrt{2}} \bm{A} - \frac{1}{\sqrt{2}} \bm{I}, \quad 
  \bm{H} = \frac{1}{\sqrt{2}} \bm{C} - \frac{1}{\sqrt{2}} \bm{G}, \nonumber \\
  \bm{C} &= R(2\theta_y) \bm{B}, \quad 
  \bm{G} = R(2\theta_x) \bm{D}, \quad 
  \bm{I} = r_\mathrm{s} R(2\theta_\mathrm{s}) \bm{H}, \nonumber \\
  \bm{b} &= \frac{1}{\sqrt{2}} \bm{a} + \frac{1}{\sqrt{2}} \bm{i}, \quad   
  \bm{d} = \frac{1}{\sqrt{2}} \bm{a} - \frac{1}{\sqrt{2}} \bm{i}, \quad
  \bm{h} = \frac{1}{\sqrt{2}} \bm{c} - \frac{1}{\sqrt{2}} \bm{g},  \\
  \bm{c} &= \exp(2i\alpha_y) R(2\theta_y) \bm{b}, \quad 
  \bm{e} = \exp(i\alpha_x) R(\theta_x) \bm{d}, \nonumber \\
  \bm{g} &= \exp(i\alpha_x) R(\theta_x) \bm{f}, \quad 
  \bm{i} = r_\mathrm{s} \exp(2i\alpha_\mathrm{s}) R(2\theta_\mathrm{s}) \bm{h}\nonumber,
\end{align}
where \(\bm{a}\) to \(\bm{i}\) represent the fluctuations in the electric fields \(\bm{A}\) to \(\bm{I}\), and \(R(\theta)\) is the rotation matrix with angle \(\theta\). 
\(\theta_x\), \(\theta_y\), and \(\theta_\mathrm{s}\) denote the phase shifts in each optical path.
Here, the arm mirrors of the Michelson interferometer are assumed to be perfectly reflecting.
\(\alpha_x\), \(\alpha_y\), and \(\alpha_\mathrm{s}\) represent the phase delays imparted to the sidebands.

Including the ponderomotive squeezing effect due to mirror displacement~\cite{pmsqz}, we obtain
\begin{equation}
  \bm{f} = \bm{e} + 2k_0 \bm{E}_{\perp}\,\delta x,\label{eq:pms}
\end{equation}
where \(k_0 = \Omega_0/c\) is the wavenumber of the carrier light, where \(\Omega_0\) is the angular frequency and \(c\) is the speed of light. 
\(\bm{E}_\perp\) is a vector perpendicular to \(\bm{E}\), defined as \(\bm{E}_\perp = R(90^\circ)\bm{E}\), and \(\delta x\) denotes the displacement of the suspended mirror.

The equation of motion describing the suspended mirror dynamics in the frequency domain ($\omega$) is given by \cite{Corbittphd}:
\begin{align}
  -m\omega^2 x &= F_0 + 2\hbar k_0\,\bm{E} \cdot \bm{e} \nonumber\\
              &= F_0 - K_\mathrm{os}\, \delta x,
\end{align}
where \(F_0\) is the ambient force acting on the mirror and \(K_\mathrm{os}\) denotes the optical spring constant. Assuming a dark port condition for the Michelson interferometer \((\theta_x - \theta_y = n\pi,\; n \in \mathbb{Z})\) and negligible asymmetry between the interferometer arms \((\alpha_x \sim \alpha_y)\), the optical spring constant can be derived by solving Eqs.~(1)–(3):
\begin{align}
  K_\mathrm{os}(\alpha) &= \frac{P_0 \, \Omega_0}{c^2} \frac{2\sin 2\phi_\mathrm{s}}{\frac{1}{r_\mathrm{s}} e^{2i\alpha} + r_\mathrm{s} e^{-2i\alpha} - 2\cos 2\phi_\mathrm{s}},
\end{align}
where \(P_0\) is the input laser power at the beamsplitter, \(r_\mathrm{s}\) the amplitude reflectance of the SRM, and \(\phi_\mathrm{s} = \theta_\mathrm{s} + (\theta_x + \theta_y)/2\) the detuning angle of the signal-recycling cavity. 
\(\alpha\) is the phase delay accumulated during a half round trip in the signal-recycling cavity:
\begin{align}
    \alpha &=\alpha_\mathrm{s}+\frac{\alpha_x+\alpha_y}{2}\\
    &= -\omega\frac{L_\mathrm{s} + (L_x+L_y)/2}{c},
\end{align}
where \(L_\mathrm{s}\), \(L_x\), and \(L_y\) denote the distances from the beam splitter to the SRM, the suspended mirror, and the end mirror on the opposite arm, respectively.  
Consequently, the optical spring frequency \(\omega_\mathrm{os}\) and damping factor \(\gamma_\mathrm{os}\) are given by:
\begin{align}
  K_\mathrm{os}&\simeq\frac{P_0 \, \Omega_0}{ c^2} \frac{2\sin 2\phi_\mathrm{s}}{\frac{1}{r_\mathrm{s}}+r_\mathrm{s} - 2\cos 2\phi_\mathrm{s}}\left(1+i\frac{T_\mathrm{s} \alpha}{1+r_\mathrm{s}^2-2r_\mathrm{s}\cos{2\phi_\mathrm{s}}}\right)\quad (|\alpha|\ll 1)\\
  &=k_\mathrm{os}+i\Gamma_\mathrm{os}\omega\\
  \omega_\mathrm{os}&=\sqrt{\frac{k_\mathrm{os}}{m}},\ \quad\gamma_\mathrm{os}=\frac{\Gamma_\mathrm{os}}{2m},
  \label{eq:omegaos}
\end{align}
where $T_\mathrm{s} = 1 - r_\mathrm{s}^2$ denotes the intensity transmissivity of the SRM.

The optomechanical properties described by Eq.~(\ref{eq:pms}) characterize the dynamic response of the suspended mirror. Consequently, the optical spring constant can be extracted from the frequency-dependent response of the system.

\subsection{Comparison of optical spring constants in various interferometer configurations}
\begin{figure}[htbp]
    \centering
    \includegraphics[width=0.5\linewidth]{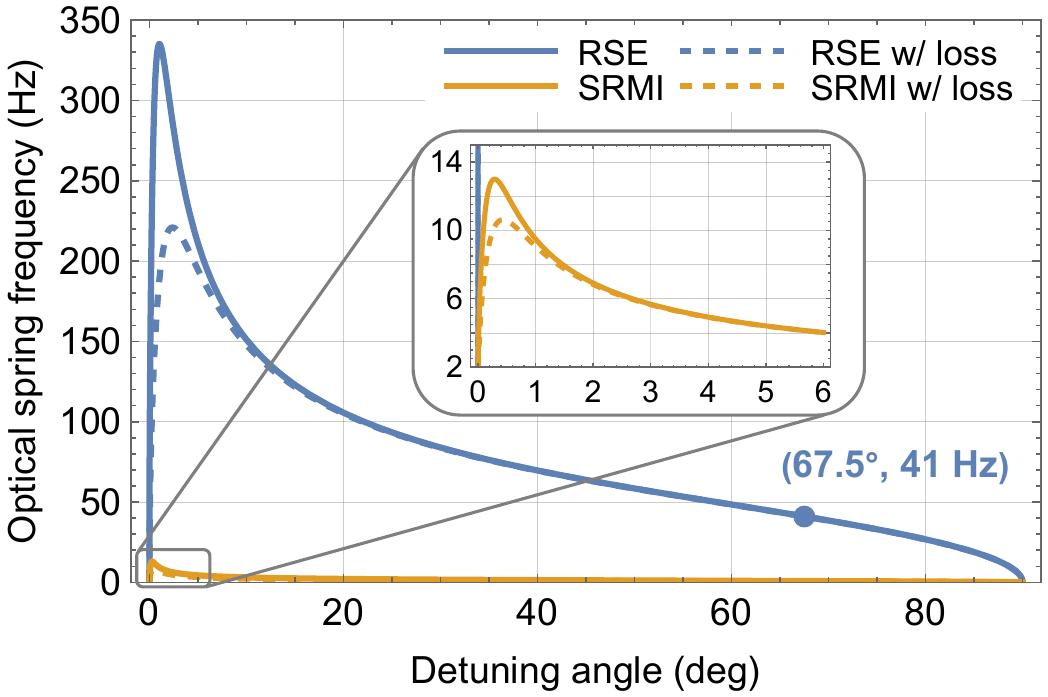}
    \caption{Estimated optical spring frequencies for the RSE and SRMI configurations. Solid lines correspond to optical and mechanical parameters taken from the Caltech 40\,m prototype~\cite{osrse} for RSE and from GEO600~\cite{GEOOS} for SRMI. Dashed lines depict the influence of signal-recycling cavity losses: 9\% for RSE and 1\% for SRMI. The 9\% value reflects experimentally observed loss during optical spring measurements \cite{RSE_OS_attempt}, whereas the 1\% value for SRMI is an estimate based on expected signal-recycling cavity losses, including Michelson contrast defects.}
    \label{fig:OShikaku}
\end{figure}
Fig.~\ref{fig:OShikaku} presents a comparison of the estimated optical spring frequencies, $f_\mathrm{os}=\omega_\mathrm{os}/(2\pi)$, for RSE and SRMI configurations. Parameters used for the RSE configuration correspond to the experimental conditions of the Caltech 40\,m prototype conducted in 2006 \cite{osrse}: arm length of \(L_\mathrm{arm} = 38.55\)\,m, laser power at the beam splitter \(P_\mathrm{BS} = 4.2\)\,W, test mass \(m = 1.276\)\,kg, SRM reflectivity \(R_\mathrm{s} = 93\%\), and arm cavity finesse \(\mathcal{F}_\mathrm{arm} = 1200\) Under these conditions, the optical spring frequency shift was observed to be 41\,Hz at a detuning angle of 67.5$^\circ$. 
Similarly, the SRMI parameters are based on the GEO600 experiment conducted in 2003 \cite{GEOOS}, with a folded arm length of \(L_\mathrm{arm} = 1200\)\,m, laser power at the beam splitter \(P_\mathrm{BS} = 1.2\)\,kW, test mass \(m = 5.6\)\,kg, and SRM reflectivity \(R_\mathrm{s} = 98\%\).

While the Caltech 40\,m prototype employed Fabry–Perot cavities in each arm, GEO600 utilized folded arm configurations containing two mirrors per arm without dedicated arm cavities.
Accordingly, the optical spring constants for these configurations can be expressed as follows:\begin{align}
    k_{\mathrm{os}, \mathrm{40\,\mathrm{m}}} 
    &= 2 \times \left(\frac{2}{\pi}\mathcal{F}_\mathrm{arm}\right)^2\, k_{\mathrm{os}}\\
    &=\frac{8\mathcal{F}_\mathrm{arm}}{\pi}\frac{P_\mathrm{arm} \, \Omega_0}{ c^2} \frac{2\sin 2\phi_\mathrm{s}}{\frac{1}{r_\mathrm{s}}+r_\mathrm{s} - 2\cos 2\phi_\mathrm{s}}\\
    k_{\mathrm{os}, \mathrm{GEO}} 
    &= 2 \times 5\, k_{\mathrm{os}}\\
    &=20\frac{P_\mathrm{arm} \, \Omega_0}{ c^2} \frac{2\sin 2\phi_\mathrm{s}}{\frac{1}{r_\mathrm{s}}+r_\mathrm{s} - 2\cos 2\phi_\mathrm{s}},
\end{align}
where $P_\mathrm{arm}$ is the circulating power in each arm.
In RSE, the phase flips upon reflection from the input test masses.
Therefore, the effective detuning angle is given by
$\phi_\mathrm{s} = \theta_\mathrm{s} + \frac{\theta_x + \theta_y}{2} + \frac{\pi}{2}$.
Arm cavities enhance the optical response to mirror displacement and scale the optical spring constant by a factor of $\mathcal{F}_\mathrm{arm}$ for a given circulating power. 
Consequently, RSE interferometers typically exhibit stronger optical rigidity than SRMI configurations under equivalent conditions.
In contrast, SRMI configurations are free from contrast-defect–induced leakage associated with arm cavities, leading to reduced optical losses. This allows for higher finesse in the SRC, yielding optical spring resonance frequencies comparable to those in RSE configurations.
Furthermore, the lower loss in SRMI makes intracavity signal amplification more effective. 
Additionally, the absence of arm cavities leads to a higher cavity pole frequency, which together makes SRMI particularly well suited for enhancing optical spring effects via intracavity amplification.

To observe strong optical springs in SRMI, the detuning phase of the cavity should be tightly controlled near the peak of the optical spring resonance. 
If the detuning is not precisely controlled, the resonance frequency drops significantly, making experimental observation difficult. 
Moreover, since optical loss in the SRC has a greater impact near the resonance peak, reducing such losses is also crucial for observing the spring in that region.
Achieving optical spring frequencies similar to those observed in RSE interferometers at large detuning—where the influence of cavity losses and detuning fluctuations is mitigated—would require megawatt-level laser power at the beam splitter. However, thermal lensing effects \cite{GEOTL} constrain realistic input powers to the kilowatt scale, posing practical limitations on experimental implementation.
These considerations highlight the fundamental experimental challenges of realizing measurable optical springs in SRMI configurations, in contrast to the relatively more favorable conditions of RSE interferometers.

\section{Experiment}\label{sec:experiment}
\subsection{Setup}
\addtocounter{figure}{-2}
\begin{figure}[htbp]
    \centering
    \begin{subfigure}
        \centering
        \captionsetup{labelformat=empty}
        
        \caption{(a)}\includegraphics[width=0.6\linewidth]{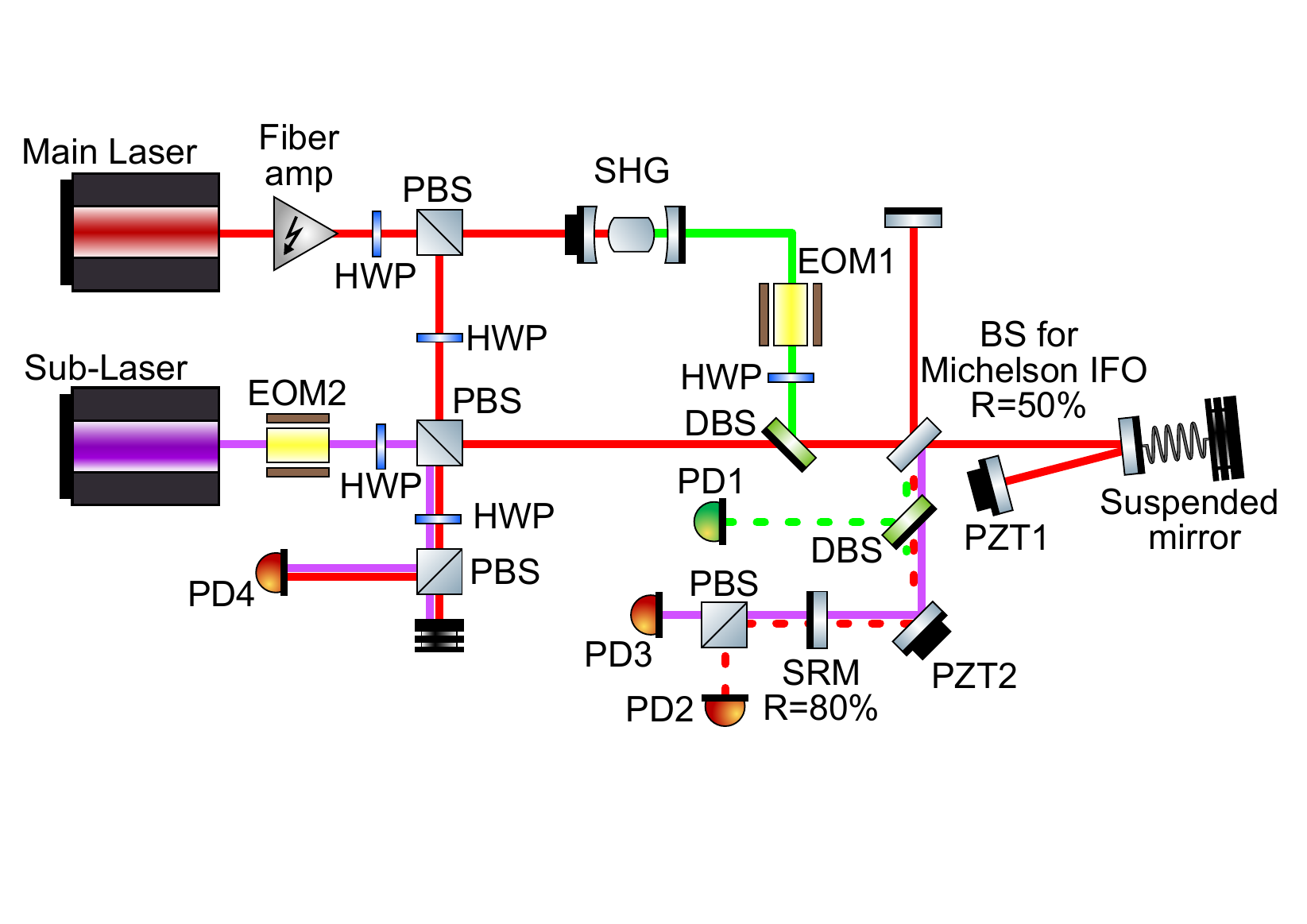}
        
    \end{subfigure}
    \begin{subfigure}
        \centering
        \captionsetup{labelformat=empty}
        \renewcommand{\thesubfigure}{}
        \caption{(b)}
        \includegraphics[width=0.4\linewidth]{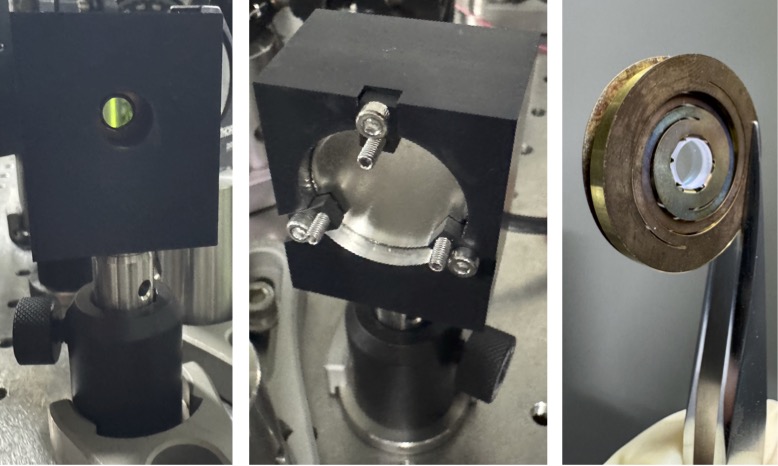}
    \end{subfigure}
    \caption{(a) Experimental setup of the SRMI, featuring a suspended mirror in one arm. A sub-laser and a green beam —generated via second-harmonic generation (SHG) of the main laser—are used for system control. (b) Photograph of the 0.2 g suspended mirror mounted within a structure containing an aperture (left). A neodymium magnet is positioned at the back to provide eddy current damping (center). The mirror is suspended using beryllium-copper double-spiral springs clamped at both sides (right).}
    \label{Fig:exp}
\end{figure}

The experimental arrangement employed in this study is illustrated in Fig.~\ref{Fig:exp}~(a).
A continuous-wave Nd:YAG laser labeled as the "Main Laser," operating at a wavelength of 1064\,nm, served as the primary carrier field. 
This beam was subsequently amplified up to 10\,W by a fiber amplifier and injected as s-polarized light into a 50:50 beam splitter (BS). 
The polarization and beam paths were controlled using a series of half-wave plates (HWPs), polarizing beam splitters (PBSs), and dichroic beam splitters (DBSs).

Within the Michelson interferometer, one arm consisted entirely of fixed mirrors, while the other arm was folded and included a suspended mirror and a piezo-actuated mirror (PZT1).
The fixed arm had an optical path length of 85\,cm, whereas the suspended mirror arm was 65\,cm long. As shown in Fig.~\ref{Fig:exp}~(b), the suspended mirror, weighing approximately 0.2\,g, was positioned at the folding point of the interferometer arm. 
This configuration effectively doubles the electric field amplitude at the folding mirror surface, resulting in a fourfold increase in radiation pressure and, consequently, in the optical spring constant.
The mirror was supported by a compliant suspension system consisting of two beryllium-copper double-spiral springs, each creating flexural hinge points oriented orthogonally on opposite sides of the mirror. This dual-sided suspension arrangement provided enhanced rigidity in both pitch and yaw degrees of freedom.
The mechanical resonance frequency of the mirror suspension system was measured to be $f_\mathrm{m} = 28.73 \pm 0.05$\,Hz.  
The quality factor was deliberately reduced to $Q = 24 \pm 2$ through the application of eddy current damping, achieved by positioning a neodymium magnet behind the suspension.
The beams returning from both interferometer arms are recombined at the beam splitter and directed toward the antisymmetric (AS) port, where they are reflected by the SRM.
This configuration forms an SRC between the beam splitter and the SRM.

To control the differential arm length of the Michelson interferometer, we employed a 1\,mW green beam at 532\,nm, generated through SHG from the main 1064\,nm laser. This beam was modulated at 50.5\,MHz using an electro-optic modulator (EOM1), combined with the main carrier at the DBS, and injected into the interferometer. 
Owing to the 20\,cm Schnupp asymmetry, a portion of the green sidebands was transmitted to the AS port even when the interferometer was locked on a dark fringe \cite{Hild_2009}. This transmitted component was separated using an additional DBS and detected by PD1, then demodulated to provide the error signal for arm length stabilization, which was fed back to PZT1. 
Because the green beam does not enter the SRC, the resulting error signal remains unaffected by SRC detuning, enabling robust and detuning-independent control of the Michelson differential arm length (see also Table~\ref{tab:sensing_matrix}).

A secondary laser source (subcarrier), operating at 1064\,nm with an output power of 200\,mW, was employed to stabilize the SRC length. 
The subcarrier was frequency-modulated at 15\,MHz using the EOM2 and combined with the main carrier beam via a PBS. 
Although the subcarrier and carrier beams have nearly identical wavelengths, they were phase-locked with a tunable frequency offset, allowing precise control of SRC detuning, as described below.
Owing to the intentional frequency offset and the interferometer’s asymmetry, a portion of the subcarrier beam exits the AS port even when the Michelson interferometer is locked on a dark fringe. 
At the AS port, the modulation sidebands of the subcarrier beat with its transmitted carrier component and generate an error signal at 15\,MHz. 
This signal is demodulated by photodetector PD3 to produce the SRC length error signal, which is fed back to PZT2 for stabilization.

The SRC detuning was precisely controlled by adjusting the beat frequency $f_\mathrm{beat}$ between the carrier and subcarrier beams, which were phase-locked using a phase-locked loop (PLL) \cite{PLL}. 
Photodetector PD4 detected the beat frequency signal and provided feedback to stabilize the subcarrier frequency. 
A combination of an HWP and PBS was placed before PD4 to ensure optimal interference between the orthogonally polarized carrier and subcarrier beams. 
At a beat frequency of $f_\mathrm{beat} = 90.5$\,MHz, the SRC length was resonant for the carrier, corresponding to zero detuning. 
A frequency shift of 81\,MHz—equal to the free spectral range (FSR) of the cavity—brought the cavity back to resonance at $f_\mathrm{beat} = 171.5$\,MHz.   
Similarly, resonance was also observed at $f_\mathrm{beat} = 9.5$\,MHz, which is separated by one FSR below 90.5\, MHz.
The 9.5\,MHz offset between the subcarrier’s resonance and the exact FSR is attributed to polarization-dependent phase shifts introduced by the cavity optics. 
Because the subcarrier is orthogonally polarized relative to the main carrier, dielectric coatings and other optical components impart slightly different phase shifts, thereby altering the resonance conditions for the subcarrier field.

\begin{table*}[htbp]
    \centering
    \caption{Sensing matrix as a function of detuning angle. Each value represents the normalized sensitivity of the error signals to perturbations in the Michelson arm length \(L_\mathrm{x}\) and the signal-recycling cavity length \(L_{\mathrm{s}}\). The matrix elements are derived by modeling the error signals from experimental parameters and computing their derivatives with respect to the corresponding longitudinal degrees of freedom.}
    \label{tab:sensing_matrix}
    \resizebox{\textwidth}{!}{%
    \begin{tabular}{c lcccc}
         \shortstack{Detuning angle (deg)}&  & 
        \shortstack{AS Carrier\\ \scriptsize (81\,MHz, I-phase)} & 
        \shortstack{REFL Carrier\\\scriptsize(81\,MHz, I-phase)} & 
        \shortstack{AS Green\\\scriptsize(50.5\,MHz, I-phase)} & 
        \shortstack{AS Sub\\\scriptsize(15\,MHz, Q-phase)} \\
        \hline
        \multirow{2}{*}{0°}  & $\delta L_\mathrm{MI}$ & \textbf{1.00}& \textbf{1.00} & \textbf{1.00} & 0.49 \\
         & $\delta L_{\mathrm{s}}$ & 0.00 & 1.38 & 0.00 & \textbf{1.00} \\
        \hline
        \multirow{2}{*}{10°} & $\delta L_\mathrm{MI}$ & 0.37 & 0.22 & \textbf{1.00} & 0.48 \\
        & $\delta L_{\mathrm{s}}$ & 0.00 & 0.02 & 0.00 & 0.98 \\
        \hline
        \multirow{2}{*}{20°} & $\delta L_\mathrm{MI}$ & 0.13 & 0.03 & \textbf{1.00} & 0.47 \\
        & $\delta L_{\mathrm{s}}$ & 0.00 & 0.17 & 0.00 & 0.96 \\
        \hline \hline
        Scaling factor $\mathrm{(W/m)}$ & & $7.55\times10^7 $&$9.55\times10^7 $&$1.82\times10^3 $&$1.61\times10^6 $
    \end{tabular}
    }
\end{table*}

Table~\ref{tab:sensing_matrix} summarizes the calculated sensing matrix for error signals used to stabilize the differential arm length of the Michelson interferometer (MICH) and the length of the signal-recycling cavity (SRCL).  
For the "AS Carrier" and "REFL Carrier" signals, the listed values correspond to conditions in which MICH stabilization relies on sidebands imposed on the carrier beam. The carrier modulation frequency was set to 81\,MHz, equal to the SRC’s FSR, ensuring maximum sensitivity in the corresponding error signals. Modulation frequencies for other cases were determined experimentally to optimize sensing performance. In the table, "AS" denotes demodulated signals detected at the antisymmetric transmission port of the SRMI, while "REFL" indicates signals detected at the reflection port. 
"AS Green" refers specifically to error signals obtained by demodulating the green beam, employed for MICH control. "AS Sub" corresponds to the demodulated subcarrier signals detected at the AS port, utilized for controlling the SRCL.

As observed in Table~\ref{tab:sensing_matrix}, when employing carrier sidebands for MICH stabilization, both "AS Carrier" and "REFL Carrier" error signals exhibit substantial variations in control gain as SRC detuning changes, thus potentially impairing control stability. Furthermore, "REFL Carrier" experiences significant coupling from SRCL fluctuations, resulting in offsets in the error signal and undesired shifts of the Michelson fringe condition as SRC detuning varies. Although the SRCL error signal similarly exhibits some crosstalk from MICH variations, this coupling is not problematic since MICH is maintained at a fixed operating condition corresponding to a dark fringe.



To reduce crosstalk between MICH and SRCL signals, avoid reductions in control gain due to detuning shifts, and achieve enhanced detuning tunability, employing the second harmonic (green beam) of the carrier for MICH control proves highly advantageous.
Therefore, we adopted a control scheme in which the Michelson differential length is stabilized using the green beam at the AS port and a piezo actuator in the arm (PZT1), while the SRC length is controlled by the subcarrier beam at the AS port and a piezo actuator in the SRC (PZT2).

\subsection{Control}
\addtocounter{figure}{-2}
\begin{figure}[htbp]
    \centering
    \begin{subfigure}
        \centering
        \captionsetup{labelformat=empty}
        
        \caption{(a)}\includegraphics[width=0.75
        \linewidth]{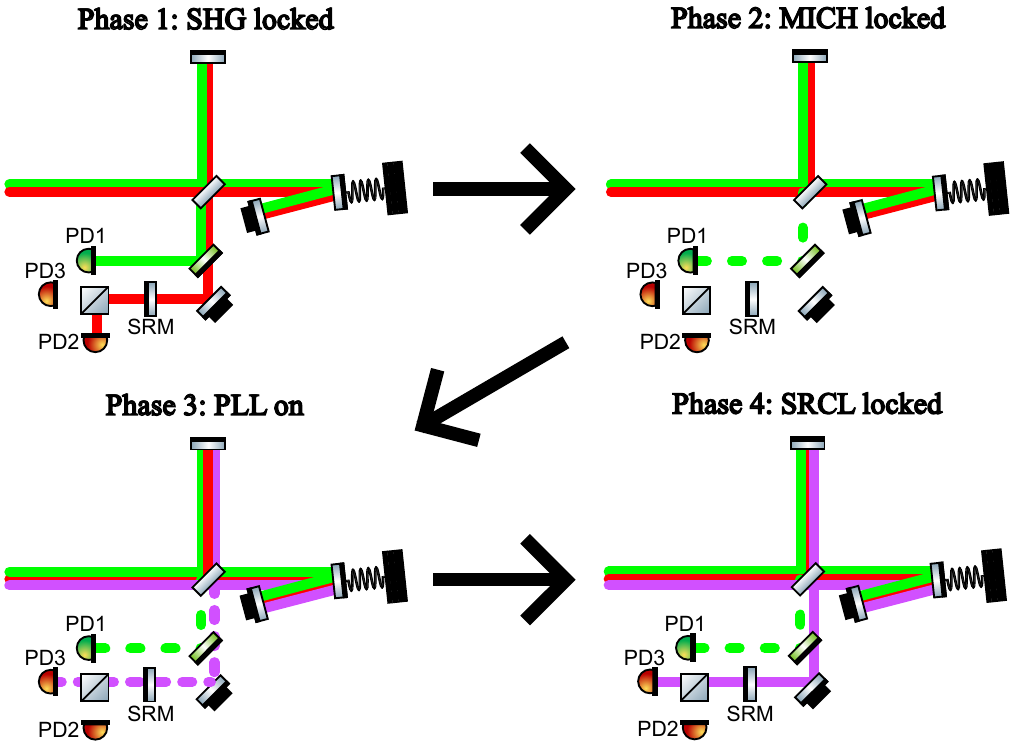}
        
    \end{subfigure}
    \begin{subfigure}
        \centering
        \captionsetup{labelformat=empty}
        \renewcommand{\thesubfigure}{}
        \caption{(b)}
        \includegraphics[width=0.75\linewidth]{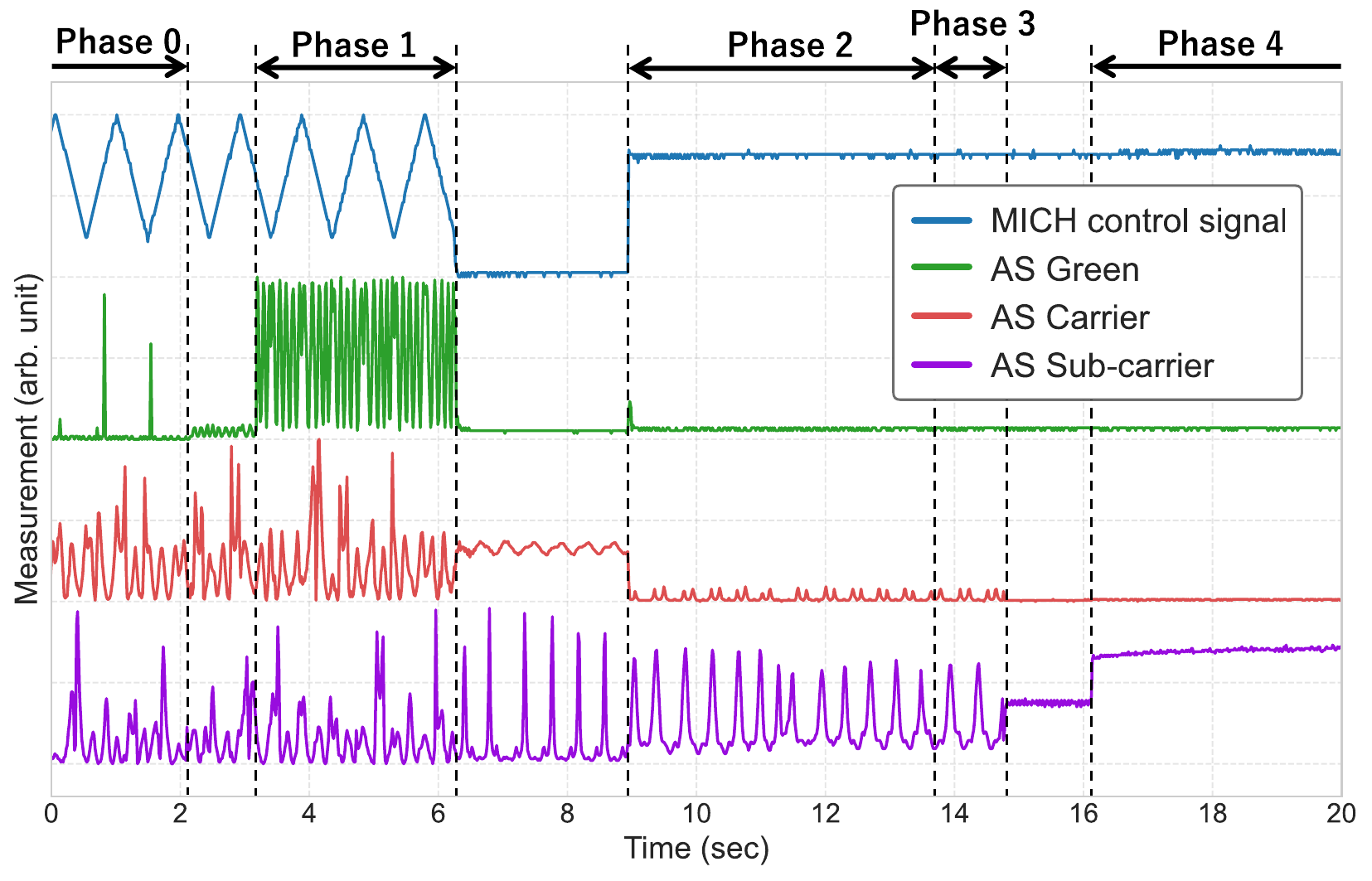}
    \end{subfigure}
    \caption{(a) Lock acquisition sequence from Phase 0 (all degrees of freedom unlocked) to Phase 4 (full lock). PD2 monitors carrier power at the AS port to confirm the Michelson fringe condition.  
    (b) Time series of the MICH control signal and DC outputs of the green, carrier, and subcarrier beams at the AS port.  
    DC locking was applied in Phases 0–1 and 3–4 to bring the system near the linear regime of the error signal.  
    An offset correction was applied between Phases 1 and 2 when MICH was mislocked to the bright fringe.}
    \label{fig:lockaq}
\end{figure}

The experiment involved active control of four distinct degrees of freedom: SHG, MICH, PLL, and SRCL.  
Fig.~\ref{fig:lockaq}~(a) depicts the sequential lock acquisition process, which comprises five phases, from Phase 0 (all degrees of freedom unlocked) to Phase 4 (full lock).  
To reliably guide the system to its intended operating point, DC locking was employed during Phases 0–1 and 3–4, bringing the system close to the linear response region of the error signal.
This approach ensured that each degree of freedom reached its optimal operating point before full feedback control was engaged.  
During the transition from Phase 1 to Phase 2, the Michelson interferometer was occasionally mislocked to the bright fringe. In such cases, an offset correction was applied, followed by re-locking to the desired dark fringe operating point.

Except for the PLL, all control loops were implemented digitally using a STEMLab 125-14 single-board computer (Red Pitaya) in conjunction with the PyRPL software package \cite{PyRPL}.  
PyRPL provides radio-frequency modulation and demodulation functionalities optimized for optical systems.  
However, the PLL required demodulation near 1\,FSR ($\sim$80\,MHz), which exceeds the Nyquist frequency (62.5\,MHz) of the STEMLab 125-14, which samples at 125\,MHz.  
To overcome this limitation, the PLL error signal was generated using a high-speed commercial phase-frequency detector (HMC439QS16G) and a high-frequency function generator. The control loop was closed via an analog low-pass filter.

Fig.~\ref{fig:lockaq}~(b) presents the time series of control signals recorded at the AS port during the lock acquisition process, along with the MICH control signal.  
For clarity, the PLL lock is depicted as occurring just prior to SRCL locking; however, in practice, the subcarrier beam was injected continuously, and the PLL typically stabilized earlier in the sequence.  
Photodetector PD2 monitored the carrier power at the AS port throughout the procedure to verify the Michelson dark fringe condition.

\subsection{Measurement}
To evaluate the optomechanical response of the interferometer system, the closed-loop transfer function \(G_\mathrm{CL}\) was measured. This measurement involved applying a frequency-swept signal to the PZT1 controlling the Michelson interferometer length and measuring the ratio of the generated control signal to the applied drive signal. The closed-loop transfer function is defined as 
\begin{equation}
  G_\mathrm{CL} = \frac{1}{1 - G_\mathrm{SRMI} \, G_\mathrm{opt}},
\end{equation}
where \(G_\mathrm{SRMI}\) represents the open-loop transfer function of the interferometer control system excluding optomechanical interactions. The optomechanical response \(G_\mathrm{opt}\) is given by \cite{cripephd}:
\begin{equation}
  G_\mathrm{opt}(\omega) = \frac{\omega_\mathrm{m}^2 - \omega^2 + 2i\gamma_\mathrm{m}\omega}
  {\omega_\mathrm{m}^2 + \omega_\mathrm{os}^2 - \omega^2 + 2i\gamma_\mathrm{m}\omega + 2i\gamma_\mathrm{os}\omega},
  \label{eq:xf}
\end{equation}
where \(\omega_\mathrm{os}\) and \(\gamma_\mathrm{os}\) denote the optical spring frequency and damping factor, respectively, and \(\omega_\mathrm{m}\) and \(\gamma_\mathrm{m}\) are the mechanical resonance frequency and mechanical damping factor of the suspension.

Transfer-function measurements were conducted at various detuning angles to characterize the behavior of the optical spring. The detuning angle was adjusted by varying the reference frequency of the PLL, which controls the beat frequency between the carrier and subcarrier beams. These measurements were performed while maintaining stable control of all experimental degrees of freedom.

\section{Results}\label{sec:results}
\addtocounter{figure}{-2}
\begin{figure}[htbp]
    \centering
    \begin{subfigure}
        \centering
        \captionsetup{labelformat=empty}
        \caption{(a)}
        \includegraphics[width=.765\columnwidth]{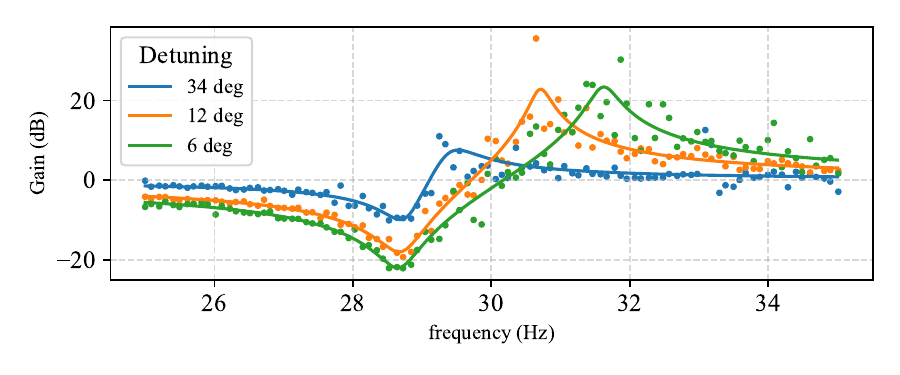}
    \end{subfigure}
    \begin{subfigure}
        \centering
        \captionsetup{labelformat=empty}
        \caption{(b)}
        \includegraphics[width=.75\columnwidth]{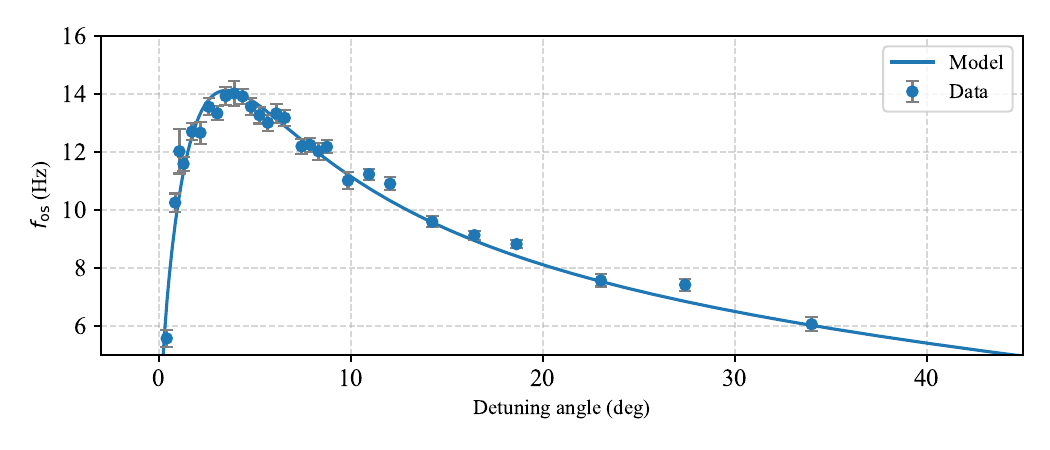}
    \end{subfigure}
    \caption{
    (a) 
    Frequency response of the optomechanical gain $|G_\mathrm{opt}(\omega)|$.  
    Data points represent experimental measurements; solid lines are the corresponding fitted curves.
    (b) 
    Data points show the measured optical spring frequencies $f_\mathrm{os} = {\omega_\mathrm{os}}/{2\pi}$ as a function of detuning angle, while the solid line indicates the theoretical prediction. Error bars denote the standard error estimated from the measured transfer function.
}
    \label{Fig:res}
\end{figure}
\noindent
The closed-loop transfer function of the system, $G_\mathrm{CL}(\omega)$, was measured across a range of detuning angles from 0.4$^\circ$ to 40.6$^\circ$. 
The gain response was subsequently converted into the optomechanical response $|G_\mathrm{opt}(\omega)|$, as shown in Fig.~\ref{Fig:res}~(a). A pronounced dip at approximately 28.7\,Hz corresponds to the mechanical resonance frequency of the suspended mirror, $f_\mathrm{m} = \omega_\mathrm{m}/(2\pi)$. The resonant peaks at higher frequencies represent the optical spring resonance, which arises from the coupling between optical and mechanical degrees of freedom and occurs at an effective frequency $f_\mathrm{eff} = \sqrt{\omega_\mathrm{m}^2 + \omega_\mathrm{os}^2}/(2\pi)$. The measured transfer function was fitted using Eq.~(\ref{eq:xf}) as a complex function to extract the system parameters.

The optical spring frequency, \(f_\mathrm{os}\) was extracted from the fitted response and plotted as a function of detuning angle in Fig.~\ref{Fig:res}~(b). 
In the fitting procedure, two free parameters were introduced: the effective mass \(m_{\mathrm{eff}}\) and the round-trip intensity loss \(\epsilon\). The parameter \(m_{\mathrm{eff}}\sim0.8\,\mathrm{g}\) represents the combined effect of the mirror and suspension system and was determined from the fit, while \(\epsilon \sim 2\%\) accounts for optical losses in the signal-recycling cavity. 
The theoretical model was adapted from Eq.~(\ref{eq:omegaos}) to account for the fourfold enhancement of the optical spring constant due to the folding geometry at the location of the suspended mirror. 
The modified expression for the optical spring frequency is given by  
\begin{eqnarray}
    \tilde\omega_{\mathrm{os}} = 2\sqrt{\frac{P_0 \, \Omega_0}{m_{\mathrm{eff}} c^2}\frac{2\sin{2\phi_\mathrm{s}}}{\frac{1}{r_\mathrm{s}\sqrt{1-\epsilon}} + r_\mathrm{s}\sqrt{1-\epsilon} - 2\cos{2\phi_\mathrm{s}}}},\label{omegaos}
\end{eqnarray}
where the input laser power \(P_0\) at the Michelson interferometer's beam splitter was measured to be approximately 7.8\,W.

As shown in Fig.~\ref{Fig:res}~(a), the position of the optical spring resonance peak shifts as a function of detuning, while the mechanical resonance dip remains fixed at the same frequency. This behavior indicates that the optomechanical coupling strength of the suspended mirror is detuning-dependent. 
As the measurements were conducted at low frequencies, the transfer function was particularly susceptible to acoustic and seismic disturbances. 
This susceptibility was most pronounced near the optical spring resonance frequency $\omega \sim \sqrt{\omega_\mathrm{m}^2 + \omega_\mathrm{os}^2}$, where the response to external disturbances is enhanced due to increased sensitivity and noise coupling, despite high open-loop gain and active feedback. 
Despite this challenge, the measured optical spring frequency as a function of detuning, shown in Fig.~\ref{Fig:res}~(b), exhibits excellent agreement with the theoretical model, yielding a coefficient of determination of $R^2 = 0.96$.
This strong correlation confirms that the observed optical spring behavior in an SRMI is consistent with predictions from optomechanical cavity theory.

\section{Outlook}\label{sec:outlook}
This study establishes a practical foundation for operating optical springs in an SRMI configuration and demonstrates its suitability for future integration of intracavity signal amplification.

The SRMI architecture is particularly well suited for implementing OPA. Unlike configurations with arm cavities, the bare SRMI minimizes optical phase delays, which would otherwise suppress optical spring enhancement. In the kilohertz regime—relevant for signals from neutron star post-merger oscillations or core-collapse supernovae—this advantage enables the SRMI-based configuration with OPA to potentially surpass the sensitivity of conventional RSE interferometers~\cite{Somiya2023}.
For example, in the case of GEO600, a circulating power of several tens of kilowatts at the beam splitter is sufficient to significantly exceed the sensitivity of current detectors. 
Furthermore, by optimizing for a narrowband response, it becomes feasible to detect kilohertz signals with even lower power levels.

Additionally, operating the signal-recycling cavity at a dark port suppresses carrier injection into the nonlinear crystal, thereby mitigating thermal loading concerns during intracavity amplification~\cite{OtabePT}. This configuration also enables selective amplification of signal sidebands while avoiding pump depletion effects, which often limit the performance of Fabry–Perot-based OPA implementations~\cite{Zhang2023}.

When OPA is incorporated into our setup, the optical spring frequency $\omega_{\mathrm{os}}$, originally given by Eq.~(\ref{omegaos}), is modified as follows~\cite{Somiya2023}:
\begin{eqnarray}
\tilde{\omega}_{\mathrm{os,OPA}} = 2
\sqrt{\frac{P_0 \Omega_0}{m_{\mathrm{eff}}c^2} \frac{(s + 1/s)\sin{2\phi_\mathrm{s}} + |s - 1/s|}{\frac{1}{r_\mathrm{s}\sqrt{1 - \epsilon}} + r_\mathrm{s}\sqrt{1 - \epsilon} - (s + 1/s)\cos{2\phi_\mathrm{s}}}},
\end{eqnarray}
where $s$ is the single-pass squeezing factor.
By compensating for losses in the signal-recycling cavity (SRC), OPA theoretically permits unbounded enhancement of the optical spring constant once the gain exceeds the oscillation threshold. In our experimental configuration, this threshold corresponds to an OPA gain of approximately 1\,dB, beyond which the optical spring frequency---and thus optical rigidity---increases rapidly.
However, this enhanced optical rigidity is accompanied by a significant increase in optical anti-damping, which can destabilize the control system. To counteract this effect, active damping techniques---such as intensity modulation and subcarrier-based damping---may be required~\cite{Corbitt2007,Singh2016,Aronson:24}.

\section{Conclusion}
We presented the first experimental observation of an optical spring in an SRMI without arm cavities. The measured optomechanical response aligns well with theoretical predictions for a conventional optomechanical cavity and remains consistent over diverse detuning angles. These findings establish a solid experimental foundation for realizing optical rigidity in simplified interferometer configurations.
Owing to its inherent compatibility with intracavity signal amplification schemes, the SRMI platform offers a promising pathway toward enhancing both broadband and high-frequency sensitivity in future gravitational-wave detectors.

In the future, we will integrate an intracavity OPA into our setup to experimentally demonstrate OPA-enhanced optical spring behavior. 
Given that increased optical rigidity can introduce significant anti-damping effects, we will also investigate active damping strategies to ensure stability of the control system.

\section*{Funding}
Japan Society for the Promotion of Science (23KJ0954, 17H02886, 23K25896);
Japan Science and Technology Agency (JPMJSC2209, JPMJCR1873, JPMJAP2320);
Foundation for the Promotion of the Open University of Japan.
\section*{Acknowledgment}
The authors thank John Winterflood at the University of Western Australia for designing the double-spiral spring,  
and the Design and Manufacturing Division of the Open Facility Center at the Institute of Science Tokyo for fabricating the Be-Cu suspension. 
We also acknowledge Satoru Takano at the Albert Einstein Institute for his valuable advice on the implementation of magnetic damping in the suspension system. 
Special thanks are extended to Pierre-Edouard Jacquet and Samuel Deléglise at Laboratoire Kastler Brossel for developing PyRPL and providing technical support for its application in this study.

\section*{Disclosures}
The authors declare no conflicts of interest.
\section*{Data Availability}
Data underlying the results presented in this paper are available from the corresponding author upon reasonable request.


\bibliography{SRMIOS}

\begin{thebibliography}{10}
\newcommand{\enquote}[1]{``#1''}

\bibitem{150914}
B.~P. Abbott, R.~Abbott, T.~D. Abbott, \emph{et~al.}, \enquote{Observation of gravitational waves from a binary black hole merger,} {\protect\JournalTitle{Phys. Rev. Lett.}} \textbf{116}, 061102 (2016).

\bibitem{alarts}
{LIGO Scientific Collaboration}, \enquote{{Gravitational-Wave Candidate Event Database},} Available at: \url{https://gracedb.ligo.org} (accessed Aug. 24, 2025).

\bibitem{BNS}
N.~Sarin and P.~D. Lasky, \enquote{The evolution of binary neutron star post-merger remnants: a review,} {\protect\JournalTitle{General Relativity and Gravitation}} \textbf{53}, 59 (2021).

\bibitem{LIGO}
{LIGO Scientific Collaboration}, J.~{Aasi}, B.~P. {Abbott}, \emph{et~al.}, \enquote{{Advanced LIGO},} {\protect\JournalTitle{Classical and Quantum Gravity}} \textbf{32}, 074001 (2015).

\bibitem{virgo}
F.~Acernese, M.~Agathos, K.~Agatsuma, \emph{et~al.}, \enquote{Advanced {Virgo}: a second-generation interferometric gravitational wave detector,} {\protect\JournalTitle{Classical and Quantum Gravity}} \textbf{32}, 024001 (2014).

\bibitem{KAGRA}
T.~Akutsu, M.~Ando, K.~Arai, \emph{et~al.}, \enquote{{KAGRA}: 2.5 generation interferometric gravitational wave detector,} {\protect\JournalTitle{Nature Astronomy}} \textbf{3}, 35–40 (2019).

\bibitem{LIGOsqz}
J.~Abadie, B.~P. Abbott, R.~Abbott, \emph{et~al.}, \enquote{A gravitational wave observatory operating beyond the quantum shot-noise limit.} {\protect\JournalTitle{Nature Physics}} \textbf{7}, 962–965 (2011).

\bibitem{LIGOfdsqz}
D.~Ganapathy, W.~Jia, M.~Nakano, \emph{et~al.}, \enquote{Broadband quantum enhancement of the ligo detectors with frequency-dependent squeezing,} {\protect\JournalTitle{Phys. Rev. X}} \textbf{13}, 041021 (2023).

\bibitem{virgofdsqz}
F.~Acernese, M.~Agathos, A.~Ain, \emph{et~al.}, \enquote{Frequency-dependent squeezed vacuum source for the advanced {Virgo} gravitational-wave detector,} {\protect\JournalTitle{Phys. Rev. Lett.}} \textbf{131}, 041403 (2023).

\bibitem{cavoptmech}
M.~Aspelmeyer, T.~J. Kippenberg, and F.~Marquardt, \enquote{Cavity optomechanics,} {\protect\JournalTitle{Rev. Mod. Phys.}} \textbf{86}, 1391--1452 (2014).

\bibitem{osshread}
B.~S. Sheard, M.~B. Gray, C.~M. Mow-Lowry, \emph{et~al.}, \enquote{Observation and characterization of an optical spring,} {\protect\JournalTitle{Phys. Rev. A}} \textbf{69}, 051801 (2004).

\bibitem{lensing}
K.~Strain, K.~Danzmann, J.~Mizuno, \emph{et~al.}, \enquote{Thermal lensing in recycling interferometric gravitational wave detectors,} {\protect\JournalTitle{Physics Letters A}} \textbf{194}, 124--132 (1994).

\bibitem{parainst}
M.~Evans, S.~Gras, P.~Fritschel, \emph{et~al.}, \enquote{Observation of parametric instability in advanced ligo,} {\protect\JournalTitle{Phys. Rev. Lett.}} \textbf{114}, 161102 (2015).

\bibitem{somiya2016}
K.~Somiya, Y.~Kataoka, J.~Kato, \emph{et~al.}, \enquote{Parametric signal amplification to create a stiff optical bar,} {\protect\JournalTitle{Physics Letters A}} \textbf{380}, 521--524 (2016).

\bibitem{korobko2017}
M.~Korobko, L.~Kleybolte, S.~Ast, \emph{et~al.}, \enquote{Beating the standard sensitivity-bandwidth limit of cavity-enhanced interferometers with internal squeezed-light generation,} {\protect\JournalTitle{Phys. Rev. Lett.}} \textbf{118}, 143601 (2017).

\bibitem{mani}
M.~Hossein-Zadeh and K.~J. Vahala, \enquote{Observation of optical spring effect in a microtoroidal optomechanical resonator,} {\protect\JournalTitle{Opt. Lett.}} \textbf{32}, 1611--1613 (2007).

\bibitem{anets}
G.~Anetsberger, O.~Arcizet, Q.~Unterreithmeier, \emph{et~al.}, \enquote{Near-field cavity optomechanics with nanomechanical oscillators,} {\protect\JournalTitle{Nature Physics}} \textbf{5}, 909--914 (2009).

\bibitem{osrse}
O.~Miyakawa, R.~Ward, R.~Adhikari, \emph{et~al.}, \enquote{Measurement of optical response of a detuned resonant sideband extraction gravitational wave detector,} {\protect\JournalTitle{Physical Review D}} \textbf{74}, 022001 (2006).

\bibitem{oscorbit}
T.~Corbitt, D.~Ottaway, E.~Innerhofer, \emph{et~al.}, \enquote{Measurement of radiation-pressure-induced optomechanical dynamics in a suspended {Fabry}-{Perot} cavity,} {\protect\JournalTitle{Phys. Rev. A}} \textbf{74}, 021802 (2006).

\bibitem{Cripe2018}
J.~Cripe, B.~Danz, B.~Lane, \emph{et~al.}, \enquote{Observation of an optical spring with a beam splitter,} {\protect\JournalTitle{Opt. Lett.}} \textbf{43}, 2193--2196 (2018).

\bibitem{otabe2024}
S.~Otabe, W.~Usukura, K.~Suzuki, \emph{et~al.}, \enquote{Kerr-enhanced optical spring,} {\protect\JournalTitle{Phys. Rev. Lett.}} \textbf{132}, 143602 (2024).

\bibitem{Liu:25}
J.~Liu, J.~Pan, C.~Blair, \emph{et~al.}, \enquote{Amplifying optical spring effect in an optical cavity with an optical parametric amplifier,} {\protect\JournalTitle{Opt. Lett.}} \textbf{50}, 2578--2581 (2025).

\bibitem{OtabePT}
S.~Otabe, K.~Komori, K.~Harada, \emph{et~al.}, \enquote{Photothermal effect in macroscopic optomechanical systems with an intracavity nonlinear optical crystal,} {\protect\JournalTitle{Opt. Express}} \textbf{30}, 42579--42593 (2022).

\bibitem{WLC}
Y.~Ma, H.~Miao, C.~Zhao, and Y.~Chen, \enquote{Quantum noise of a white-light cavity using a double-pumped gain medium,} {\protect\JournalTitle{Phys. Rev. A}} \textbf{92}, 023807 (2015).

\bibitem{GEO_OS_attempt}
S.~Hild, \enquote{Attempts to measure the optical spring in {GEO}600,} in \emph{GEO Simulation Meeting,}  (2008).

\bibitem{Adhikari2008}
R.~Adhikari, Y.~Aso, S.~Ballmer, \emph{et~al.}, \enquote{Upgrade of the 40m interferometer,} Tech. Rep. LIGO-T080074-00-R, California Institute of Technology and Massachusetts Institute of Technology (2008).

\bibitem{Mullavey:12}
A.~J. Mullavey, B.~J.~J. Slagmolen, J.~Miller, \emph{et~al.}, \enquote{Arm-length stabilisation for interferometric gravitational-wave detectors using frequency-doubled auxiliary lasers,} {\protect\JournalTitle{Opt. Express}} \textbf{20}, 81--89 (2012).

\bibitem{IOrel}
T.~Corbitt, Y.~Chen, and N.~Mavalvala, \enquote{Mathematical framework for simulation of quantum fields in complex interferometers using the two-photon formalism,} {\protect\JournalTitle{Phys. Rev. A}} \textbf{72}, 013818 (2005).

\bibitem{pmsqz}
H.~J. Kimble, Y.~Levin, A.~B. Matsko, \emph{et~al.}, \enquote{Conversion of conventional gravitational-wave interferometers into quantum nondemolition interferometers by modifying their input and/or output optics,} {\protect\JournalTitle{Phys. Rev. D}} \textbf{65}, 022002 (2001).

\bibitem{Corbittphd}
T.~R. Corbitt, \enquote{Quantum noise and radiation pressure effects in high power optical interferometers,} Ph.D. thesis, Georgia Institute of Technology (2001).

\bibitem{GEOOS}
J.~Harms, Y.~Chen, S.~Chelkowski, \emph{et~al.}, \enquote{Squeezed-input, optical-spring, signal-recycled gravitational-wave detectors,} {\protect\JournalTitle{Phys. Rev. D}} \textbf{68}, 042001 (2003).

\bibitem{RSE_OS_attempt}
O.~Miyakawa, \enquote{Results of 40m prototype,} in \emph{LSC meeting at LHO,}  (2006).

\bibitem{GEOTL}
S.~Nadji, H.~Wittel, N.~Mukund, \emph{et~al.}, \enquote{{GEO}600 beam splitter thermal compensation system: new design and commissioning,} {\protect\JournalTitle{arXiv:2408.02804}}  (2024).

\bibitem{Hild_2009}
S.~Hild, H.~Grote, J.~Degallaix, \emph{et~al.}, \enquote{{DC}-readout of a signal-recycled gravitational wave detector,} {\protect\JournalTitle{Classical and Quantum Gravity}} \textbf{26}, 055012 (2009).

\bibitem{PLL}
G.~Santarelli, A.~Clairon, S.~Lea, and G.~Tino, \enquote{Heterodyne optical phase-locking of extended-cavity semiconductor lasers at 9 {GHz},} {\protect\JournalTitle{Optics Communications}} \textbf{104}, 339--344 (1994).

\bibitem{PyRPL}
L.~Neuhaus, M.~Croquette, R.~Metzdorff, \emph{et~al.}, \enquote{Python red pitaya lockbox ({PyRPL}): An open source software package for digital feedback control in quantum optics experiments,} {\protect\JournalTitle{Review of Scientific Instruments}} \textbf{95}, 033003 (2024).

\bibitem{cripephd}
J.~D. Cripe, \enquote{Broadband measurement and reduction of quantum radiation pressure noise in the audio band,} Ph.D. thesis, Louisiana State University and Agricultural and Mechanical College (2018).

\bibitem{Somiya2023}
K.~Somiya, K.~Suzuki, S.~Otabe, and K.~Harada, \enquote{Intracavity signal amplification system for next-generation gravitational-wave detectors,} {\protect\JournalTitle{Phys. Rev. D}} \textbf{107}, 122005 (2023).

\bibitem{Zhang2023}
J.~Zhang, H.~Sun, H.~Guo, \emph{et~al.}, \enquote{Optical spring effect enhanced by optical parametric amplifier,} {\protect\JournalTitle{Applied Physics Letters}} \textbf{122}, 261106 (2023).

\bibitem{Corbitt2007}
T.~Corbitt, Y.~Chen, E.~Innerhofer, \emph{et~al.}, \enquote{An all-optical trap for a gram-scale mirror,} {\protect\JournalTitle{Physical Review Letters}} \textbf{98}, 150802 (2007).

\bibitem{Singh2016}
R.~Singh, G.~D. Cole, J.~Cripe, and T.~Corbitt, \enquote{Stable optical trap from a single optical field utilizing birefringence,} {\protect\JournalTitle{Physical Review Letters}} \textbf{117}, 213604 (2016).

\bibitem{Aronson:24}
S.~Aronson, R.~Pagano, T.~Cullen, \emph{et~al.}, \enquote{Optical spring tracking for enhancing quantum-limited interferometers,} {\protect\JournalTitle{Opt. Lett.}} \textbf{49}, 6980--6983 (2024).

\end{thebibliography}






\end{document}